\documentclass[aps,prd,twocolumn,a4paper]{revtex4-1}
\usepackage{ulem}
\usepackage{color}
\usepackage{graphicx}
\usepackage{hyperref}   
\usepackage{appendix}
\usepackage{epsfig}
\usepackage{epstopdf}
\usepackage{amsmath}
\usepackage{subfig}
\usepackage{comment}
\usepackage[dvipsnames]{xcolor}
\newcommand{\be}{\begin{equation}}
\newcommand{\ee}{\end{equation}}
\newcommand{\ba}{\begin{eqnarray}}
\newcommand{\ea}{\end{eqnarray}}

\begin{document}
\title{Effective description of hot QCD medium in strong magnetic field and longitudinal conductivity }
\author{Manu Kurian }
\email{manu.kurian@iitgn.ac.in}
\author{Vinod Chandra }
\email{vchandra@iitgn.ac.in}
\affiliation{Indian Institute of Technology Gandhinagar,  Gandhinagar-382355, Gujarat, India}

\begin{abstract}
Hot QCD medium effects have been studied in the effective quasi-particle description of quark-gluon plasma. This model encodes the collective excitation of gluons and quarks/anti-quarks in the thermal medium in terms of effective quarks and gluons having non-trivial energy dispersion relation. The present investigation involves the extension of the effective quasi-particle model in strong magnetic field limit. Realizing, hot QCD medium in the strong magnetic field as an effective grand canonical system in terms of the modified quark, anti-quark and gluonic degrees of freedom, the thermodynamics has been studied. Further, the Debye mass in hot QCD medium has to be sensitive to the magnetic field, and subsequently the same has been observed for the effective hot QCD coupling. As an implication, electrical conductivity (longitudinal) has been studied within an effective kinetic theory description of hot QCD in the presence of the strong magnetic field. The hot QCD equation of state (EoS), dependence entering through the effective coupling and quasi-parton distribution function, found to have a significant impact on the longitudinal electrical conductivity in strong magnetic field background.\\

\end{abstract}
\maketitle
 
\section{Introduction}
It has been observed that the strongly coupled matter, quark-gluon-plasma (QGP), produced in heavy-ion collision~\cite{STAR, Aamodt:2010pb} behaves more like a near-perfect fluid. 
Relativistic heavy-ion collisions (RHIC) might produce large electromagnetic fields~\cite{Skokov:2009qp, Zhong:2014cda}, especially in off-central asymmetric collisions. In fact, the produced magnetic field is strongest among all the known magnitudes of the magnetic fields in nature. 
Therefore, it is natural to ask whether the properties of hot and dense matter produced in RHIC, are sensitive to these fields. Importance of magnetic field in hot QCD medium has been studied in the context of heavy-ion collisions in~\cite{Tuchin:2014pka, Tuchin:2013ie, deng, Das:2016cwd} and in the early universe~\cite{Castorina:2015ava}. During off-central heavy-ion collision, magnetic field can reach to $ eB\sim (1 - 15)m_{\pi}^{2}$~\cite{Skokov:2009qp, Zhong:2014cda}. Several phenomena like magnetic catalysis~\cite{Shovkovy:2012zn}, chiral magnetic effect~\cite{Fukushima:2008xe, Huang} {\it etc.}, occur in  the QGP  in presence of magnetic field. In early cosmological stages, an extremely high magnetic field has been estimated, ranges up to $eB\sim 1$ GeV$^{2}$~\cite{Grasso:2000wj}. 
  
  In light of the above points, it can be inferred that the QCD properties in the strong magnetic background may reveal a better understanding of the QGP in RHIC. In particular, the strong magnetic field in hot QCD medium is likely to change the QCD thermodynamic~\cite{Strickland:2012vu, Menezes:2015mqa, Avancini:2017gck} and hence the macroscopic observables such as transport coefficients namely conductivity, viscosity{\it, etc.} will certainly get modified. This sets the motivation of the present investigations. At the level of hot QCD thermodynamics, we include the magnetic field by extending a recently proposed quasi-particle description of hot QCD. On the other hand, for the transport coefficients, between the two equivalent approaches to include magnetic field effects in the transport coefficients, {\it viz.}, the hard thermal loop effective theory (HTL) and  the relativistic transport theory, the latter has been chosen for our analysis, in which, the magnetic field enters through the propagator (in the collision kernel) and momentum distribution functions of the effective gluons and quarks/anti-quarks. The modifications at the level of propagator (quark) are obtained in terms of the Schwinger propagator~\cite{Schwinger:1951nm}. Here we are focusing on the modification of the distribution functions by incorporating the relativistic Landau levels in the presence of magnetic field. 
 Thermodynamic quantities like energy density, pressure, entropy density, the velocity of sound can be described using the re-defined distribution function in the strong field limit ($\mid eB\mid $ $\gg$ $T^{2}$). Furthermore, setting up an effective kinetic theory in the presence of strong magnetic field with appropriate particle distributions and non-trivial dispersions results in the theoretical estimation of the transport coefficients in the strong field background. 
As an implication, we are estimating the longitudinal electrical conductivity in strong magnetic field. 

 Notably, in the strong magnetic field, quark-antiquark annihilation and quark-antiquark pair production processes are possible~\cite{Fukushima:2011nu, Tuchin:2010gx}. Thus, along with usual 2 $\rightarrow$ 2 scattering, 1 $\rightarrow$ 2 processes are also present there. It has been realized that in the strong magnetic field, and 1 $\rightarrow$ 2 process dominates over 2 $\rightarrow$ 2 scattering while estimating the electrical conductivity~\cite{Hattori:2016cnt} of the medium. The equation of state (EoS) dependence on the electrical conductivity, diffusion coefficient and charge susceptibility in the absence of magnetic field has already been studied~\cite{Mitra:2016zdw}. 
 There are different approaches for the estimation of these transport properties. Refs.~\cite{Greif:2014oia, Puglisi:2015} describes the estimation of electrical conductivity from relativistic transport equation. The electrical conductivity (longitudinal) calculation in this work involves the formulation of effective kinetic theory in lowest Landau level (LLL) approximation by including the proper collision integral in the Boltzmann equation. Our prime focus will be on the effective running coupling constant and the partonic distribution function which allows the QCD EoS dependence on longitudinal conductivity.
   
 The fluctuations of the electromagnetic field in the heavy-ion collision experiments might play an important role. These fluctuations may affect the correlation between the magnetic field direction and reaction plane. In view of the recent work~\cite{Zakharov:2017yst}, which shows that these fluctuations are much smaller as compared to the earlier prediction on the same~\cite{Bloczynski:2012en}, the dominant contribution to the longitudinal electrical conductivity mainly comes from the uniform strong magnetic field. Therefore, while computing the conductivity, the magnetic field is taken to be spatially uniform. Note that the Debye mass ($m_{D}$), is another fundamental quantity of plasma that can measure the screening effects in the medium and might get large modifications in the presence of the magnetic field. This also needed to define an effective QCD coupling in the presence of the magnetic field which is also needed in the computation of the conductivity. Again, we follow kinetic theory approach to compute, $m_D$ and the effective QCD coupling.
In the present work, we utilize the effective fugacity quasi-particle model (EQPM), proposed by Chandra and Ravisankar~\cite{Chandra:2011en, Chandra:2007ca} and extend it for the QCD thermodynamics and the longitudinal conductivity in presence of  the strong magnetic field. Note that, the EQPM description of the transport properties of the QGP has been well investigated in several works~\cite{Chandra:2008hi, Bluhm:2011,Kapusta:2011, Kapusta:2016, Das:2012ck}. 
In this context, the hot QCD medium dependence to various transport coefficients (shear and bulk viscosities, electrical conductivity, thermal conductivity and their respective ratios) in the absence of the magnetic field is well understood in Ref.~\cite{Mitra:2016zdw}. 
 
  The paper is organized as follows. In section II, an extension of the effective quasi-particle model in the strong magnetic field is discussed along with calculation of the QCD thermodynamic quantities in the strong field limit. 
Section III describes the change in Debye screening mass ($m_{D}$) and consequently the modified effective running coupling ($\alpha_{eff}(T)$), in the presence of strong magnetic field. 
Section IV deals with the quasi-particle description of extended kinetic theory in LLL approximation and the calculation of longitudinal conductivity in leading order perturbative QCD. 
Summary and conclusions have been presented in section V.

\section{Extension of EQPM and QCD thermodynamics in strong magnetic field}
Strong magnetic background plays a vital role in the quantization of fermionic theory. The energy eigenvalues are obtained as relativistic Landau levels and well investigated in several recent works~\cite{Bhattacharya:2007vz, Bruckmann:2017pxd}.
 In Landau gauge $A_{y}=Bx$ (such that $\bf{B}$ $= B\hat{z}$), the relativistic Landau levels leads to  the energy eigenvalues as, 
\begin{equation}
E_{l}=\sqrt{m^{2}+p_{z}^{2}+2l\mid q_{f}B\mid},
\end{equation}
where $q_{f}$ is the charge of fermion and $l=0,1,2,..$ is the order of the Landau energy levels. The order of energy for higher Landau levels goes as $\sqrt{\mid q_{f}B\mid}$. In LLL approximation $T$$^{2}$$\ll$ $\mid q_{f}B\mid$, the occupation in higher levels is negligibly small. So in strong magnetic field limit, LLL $(l=0)$ 
approximation constraints the motion of a particle in the direction of magnetic field with the transverse density of states $\dfrac{\mid q_{f}B\mid}{2\pi}$. The impact of this dimensional reduction $(D$$\rightarrow $$D-2)$ have been studied for the magnetic catalysis~\cite{Gusynin:1995nb, Fukushima:2012kc}. In presence of  strong magnetic field, the integration phase factor becomes,
\begin{equation}\label{2}
\int{\dfrac{d^{3}p}{(2\pi)^{3}}}\rightarrow\dfrac{\mid q_{f}B\mid}{2\pi}\int{\dfrac{dp_{z}}{2\pi}}.
\end{equation}
We shall analyze the effects of this dimensional reduction from LLL approximation in the hot QCD thermodynamics in the subsequent sections. Before that let us proceed to discuss EQPM and its extension in strong magnetic field. 
 
\subsection{EQPM and its extension in strong magnetic field}
In the quasi-particle model, the system of interacting massless particles can be considered as the non-interacting / weakly interacting particles either with effective fugacity~\cite{Chandra:2008kz} or with effective masses~\cite{Goloviznin:1994, Peshier}. There are models that include effective masses  with Polyakov loop~\cite{D'Elia:97}, NJL and PNJL based quasi-particle models~\cite{Dumitru}, and 
self-consistent and single parameter quasiparticle models~\cite{Bannur:2006js}. Note that there are some recently proposed quasi-particle models based on the Gribov-Zwanziger (GZ) quantization, leading to a nontrivial IR-improved dispersion relation in terms of the Gribov parameter~\cite{zwig}.

The EQPM considered here, interprets the hot QCD EoS as the non-interacting quasi-gluons/quasi-quarks (quasi-partons) with effective fugacities (quasi-gluon and quasi-quark fugacities, $z_{g}$ and $z_{q}$ respectively), 
which encodes all the medium interactions. The quasi-gluon and quark/anti-quark distribution functions are  given as,
\begin{equation}\label{3}
f_{g/q}=\dfrac{z_{g/q}\exp{(-\beta E_{p})}}{1\mp z_{g/q}\exp{(-\beta E_{p})} },
\end{equation}    
where $E_{g}=\mid \vec{p}\mid \equiv p$ for gluons and  $E_{p}=\sqrt{p^{2}+m^{2}}$ for quarks/anti-quarks. We are working in units where $k_{B}=1$, $c=1$, $\hbar=1$ and hence $\beta=\dfrac{1}{T}$. The physical significance of the effective fugacity comes in the dispersion relation
\begin{equation}\label{4}
\omega_{g}=p+T^{2}\partial_{T} \ln(z_{g}),
\end{equation}
\begin{equation}\label{5}
\omega_{q}=\sqrt{p^{2}+m^{2}}+T^{2}\partial_{T} \ln(z_{q}).
\end{equation}
 The second term of the dispersion relations Eqs.~(\ref{4}) and~(\ref{5}) corresponds to the collective excitation of quasi-partons. Thus effective fugacities describe the hot QCD medium effects. Consequently, the EoS dependence of the distribution functions, enters through the effective fugacities. Both $z_{q}$ and $z_{q}$ have complicated temperature
  dependence as discussed in Ref.~\cite{Jamal:2017dqs}. Here , we consider the EQPM description of  the recent (2+1) flavor lattice QCD EoS (LEoS)~\cite{Cheng:2007jq} and 3-loop HTL perturbative (HTLpt) EOS~\cite{Haque, Andersen}. The 3-loop HTLpt EOS has recently been computed by N. Haque $et, al$. which is very close to the recent lattice results~\cite{Borsanyi, Haque:2014}. These EoSs have been carefully 
  embedded in $z_{q}$ and $z_{q}$. Since this model is valid beyond the transition temperature, the mass of light quarks is considered almost negligible. The EQPM, like effective mass models and other QP models, modify the kinetic theory definition of energy-momentum stress tensor $T^{\mu\nu}$ as provided in Ref.~\cite{Chandra:2012qq}. The effective kinetic theory to compute
   first order transport coefficients of the hot QCD medium is discussed in detail in Ref.\cite{Mitra:2017sjo}.
 
The extension of EQPM in the magnetic field involves the modification of dispersion relation by relativistic Landau levels. The quark/anti-quark distribution function can be obtained as,
\begin{equation}
f_{q}^{0}=\dfrac{z_{q}\exp{(-\beta \sqrt{p_{z}^{2}+m^{2}+2l\mid q_{f}B\mid})}}{1+ z_{q}\exp{( -\beta \sqrt{p_{z}^{2}+m^{2}+2l\mid q_{f}B\mid} )}},
\end{equation}    
where the magnetic field is taken along the z-axis direction.

Next,  the average energy can be obtained in terms of an effective Grand-Canonical partition function:
\begin{equation}
\ln Z_{q}=\nu_{q}\int{\dfrac{d^{3}p}{(2\pi)^{3}}\ln \left( 1+ z_{q}\exp{(-\beta E_{p})}\right) },
\end{equation} 
where $E_{p}=\sqrt{p_{z}^{2}+m^{2}+2l\mid q_{f}B\mid}$, following the basic thermodynamic definition,  $\varepsilon=-\dfrac{\partial\ln Z_{g/q}}{\partial\beta}$, leading to,
\begin{equation}
\varepsilon= \nu_{q}\int{\dfrac{d^{3}p}{(2\pi)^{3}}}\omega^{0}_{q}f^{0}_{q} 
\end{equation}
with $\omega_{q}^{0}=E_{p}+T^{2}\partial_{T} \ln(z_{g})$. Again, for phase-space integration dimensional reduction will work in the usual way. Under strong magnetic field, LLL approximation will work and hence the quasi-quark distribution function becomes,
\begin{equation}\label{7}
f_{q}^{0}=\dfrac{z_{q}\exp{(-\beta \sqrt{p_{z}^{2}+m^{2})}}}{1+ z_{q}\exp{(-\beta \sqrt{p_{z}^{2}+m^{2}})}}.
\end{equation}
The quark dispersion relation in the strong field limit is
\begin{equation}\label{8}
\omega_{q}^{0}=\sqrt{p_{z}^{2}+m^{2}}+T^{2}\partial_{T} \ln(z_{g}).
\end{equation}
Clearly, in the presence of magnetic field (even in LLL) the dispersion relation for quarks and anti-quarks get modified. Since gluon is charge-less, dispersion relation will remain intact. Therefore quasi-gluon distribution function in magnetic field will remain the same
($f^{0}_{g}\equiv f_{g}$).

 Next, from the extended EQPM we calculate relevant QCD thermodynamic quantities like pressure, energy density, entropy,  and velocity of sound in presence of strong magnetic field. Throughout our calculations, the QCD transition temperature is taken as $T_{c}=170$ MeV.  

\subsection{QCD thermodynamics in the presence of strong magnetic field}
 The thermodynamic quantities of our interest are number density, energy density, pressure, entropy density, and velocity of sound.
Let us start with the energy density in extended EQPM. Following definition of energy density,
\begin{equation}\label{9}
\varepsilon= \nu_{g}\int{\dfrac{d^{3}p}{(2\pi)^{3}}} \omega_{g}f_{g}+\sum_{a=q,\bar{q}} \nu_{a}\int{\dfrac{d^{3}p}{(2\pi)^{3}}}\omega_{a}f_{a} .
\end{equation}
where $v_{g}=2( N_{c}^{2}-1 )$ and $v_{q/\bar{q}}=2N_{c}N_{f}$ for $SU(N_c)$ with $N_f$ flavors.
As mentioned earlier, in presence of strong magnetic field background ($B\hat{z}$), the phase space factors as well as the quarks/anti-quarks distribution functions are modified. Following this argument and from Eqs.~(\ref{2}),~(\ref{4}),~(\ref{7}) and~(\ref{8}) the energy density becomes 
\begin{align}
\varepsilon&= \int{\dfrac{d^{3}p}{(2\pi)^{3}}}\nu_{g}\omega_{g}f_{g}\nonumber \\
&+\dfrac{\mid q_{f}B\mid}{(2\pi)}\sum_{a=q,\bar{q}} \nu_a \int_{-\infty}^{\infty}{\dfrac{dp_{z}}{(2\pi)}}\dfrac{\omega_{a}^{0}}{(z_{a}^{-1}\exp{(\beta\sqrt{p_{z}^{2}}})+1)} ,
\end{align}
which integral can be expressed in terms of $PolyLog$ functions over the fugacity parameters.
\begin{align}
\varepsilon&=\dfrac{3T^{4}}{\pi^{2}}\nu_{g}PolyLog[4,z_{g}]-\dfrac{2T^{2}}{\pi^{2}}q_{f}B\nu_qPolyLog[2,-z_{q}]\nonumber\\
&+\left( T^{2}\partial_{T}\ln z_{g}\right)\dfrac{T^{3}}{\pi^{2}}\nu_{g}PolyLog[3,z_{g}]\nonumber\\
&+\mid q_{f}B\mid\left( T^{2}\partial_{T}\ln z_{q}\right)\dfrac{T}{\pi^{2}}\nu_q\ln(1+z_{q}). 
\end{align}
Throughout the calculation, the degeneracy sum over quarks/anti-quarks includes three flavors. In the absence of magnetic field, the dimensional reduction gets switched off. Eq.~(\ref{9}) gives the usual energy density without magnetic field,

\begin{align}
\varepsilon&=\dfrac{3T^{4}}{\pi^{2}}\nu_{g}PolyLog[4,z_{g}]-\dfrac{6T^{4}}{\pi^{2}}\nu_{q}PolyLog[4,-z_{q}]\nonumber\\
&+\left( T^{2}\partial_{T}\ln z_{g}\right)\dfrac{T^{3}}{\pi^{2}}\nu_{g}PolyLog[3,z_{g}]\nonumber\\
&-\left( T^{2}\partial_{T}\ln z_{q}\right)\dfrac{T^{3}}{\pi^{2}}\nu_{q}2PolyLog[3,-z_{q}]. 
\end{align}
In strong field limit ($\mid q_{f}B\mid\gg T^{2}$), energy density with the magnetic field background is high. At very high temperature this condition gets violated so that we have to include higher Landau level corrections for the estimation of energy density.

Likewise, the particle number density is given by
\begin{equation}
n= \nu_{g} \int{\dfrac{d^{3}p}{(2\pi)^{3}}}f_{g}^{0}+\dfrac{\mid q_{f}B\mid}{(2\pi)}\sum_{a=q,\bar{q}} \nu_a \int{\dfrac{dp_{z}}{(2\pi)}}f_{a}^{0}.
\end{equation}
Thus, the modified EQPM defines the number density as
\begin{align}
n&= \nu_{g} \int_{0}^{\infty}{\dfrac{p^{2}dp}{2\pi^{2}}}\dfrac{1}{(z_{g}^{-1}\exp{(\beta p)}-1)}\nonumber\\
&+\dfrac{\mid q_{f}B\mid}{(2\pi)}\sum_{a=q,\bar{q}} \nu_a \int_{-\infty}^{\infty}{\dfrac{dp_{z}}{(2\pi)}}\dfrac{1}{(z_{q}^{-1}\exp{(\beta\sqrt{p_{z}^{2}})}+1)}\\
&=\dfrac{T^{3}}{\pi^{2}}\nu_{g}PolyLog[3,z_{g}]
+\mid q_{f}B\mid\dfrac{T}{\pi^{2}}\nu_q\ln(1+z_{q}).
\end{align}
Similarly, the behavior of  pressure $P$, entropy density $s$,  in the strong magnetic field can be obtained starting from their basic definitions,
\begin{equation}
P= \sum_{a=g,q,\bar{q}}\dfrac{1}{3} \nu_{a}\int{\dfrac{d^{3}p}{(2\pi)^{3}p_{a}^{0}}}\mid\vec{p_{a}}\mid^{2} f_{a}^{0}.
\end{equation} 

Incorporating the extended EQPM we have
\begin{align}\label{17}
P&=\dfrac{T^{4}}{\pi^{2}}\nu_{g}PolyLog[4,z_{g}]\nonumber\\
&-\mid q_{f}B\mid\dfrac{T^{2}}{\pi^{2}}\dfrac{1}{3}\nu_qPolyLog[2,-z_{q}].
\end{align}
 Entropy density can be calculated from its fundamental definition,
\begin{align}
s&=\dfrac{\varepsilon+P}{T}\\
&=\dfrac{4T^{3}}{\pi^{2}}\nu_{g}PolyLog[4,z_{g}]\nonumber\\
&-\dfrac{7\mid q_{f}B\mid T}{3\pi^{2}}\nu_qPolyLog[2,-z_{q}]\nonumber\\
&+\left( T^{2}\partial_{T}\ln z_{g}\right)\dfrac{T^{2}}{\pi^{2}}\nu_{g}PolyLog[3,z_{g}]\nonumber\\
&+\mid q_{f}B\mid\left( T^{2}\partial_{T}\ln z_{q}\right)\dfrac{\nu_q}{\pi^{2}}\ln(1+z_{q}).\label{19}
\end{align}
From pressure and energy density, enthalpy density can also be calculated in a straightforward way. Specific heat at constant pressure $c_{p}$ could also be obtained conveniently from enthalpy density.
\begin{figure*}
  \centering
  \hspace{-17mm}
\subfloat{\includegraphics[height=7.650cm,width=5.85cm]{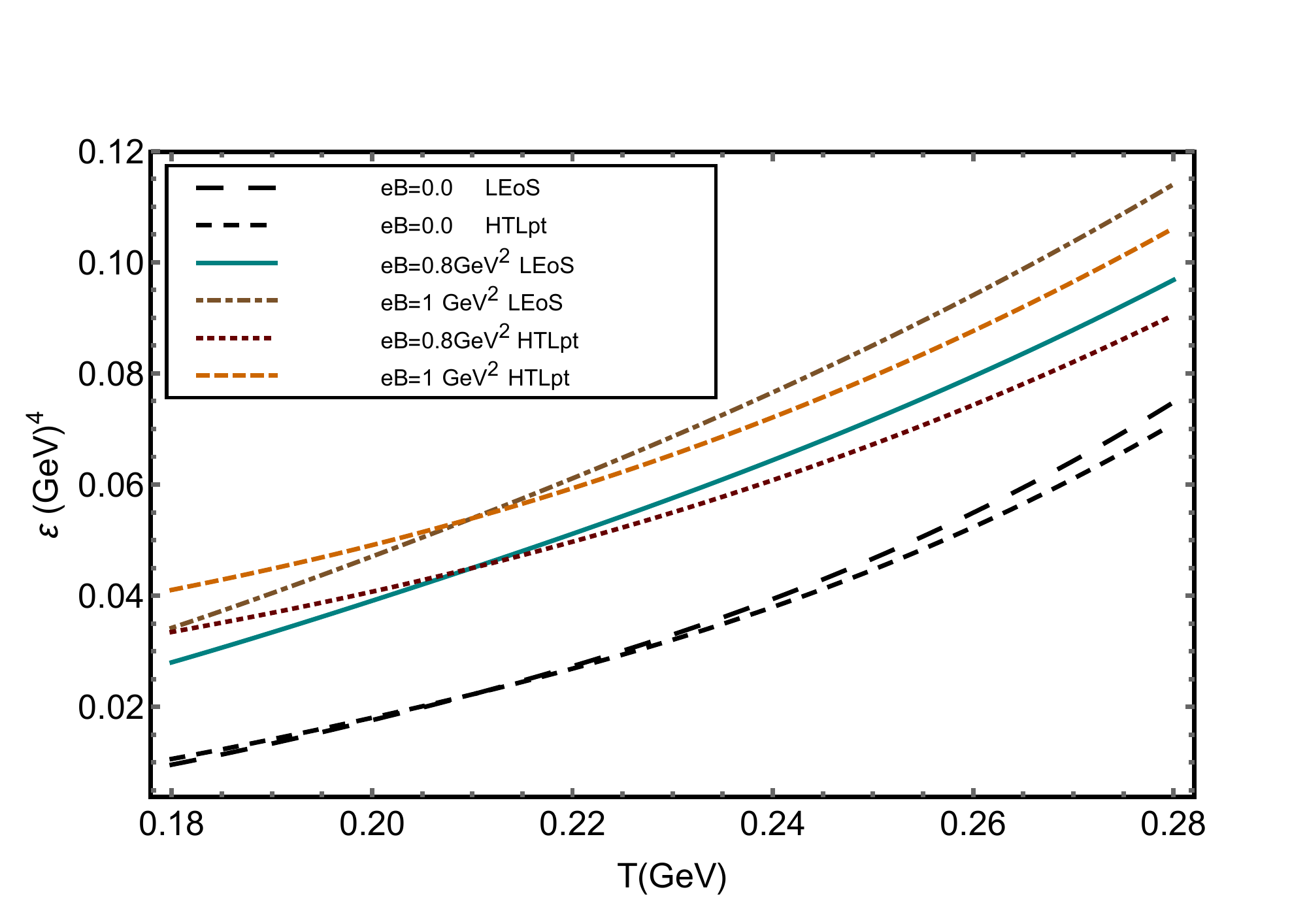}}
\hspace{-.5mm}
\subfloat{\includegraphics[height=7.50cm,width=5.80cm]{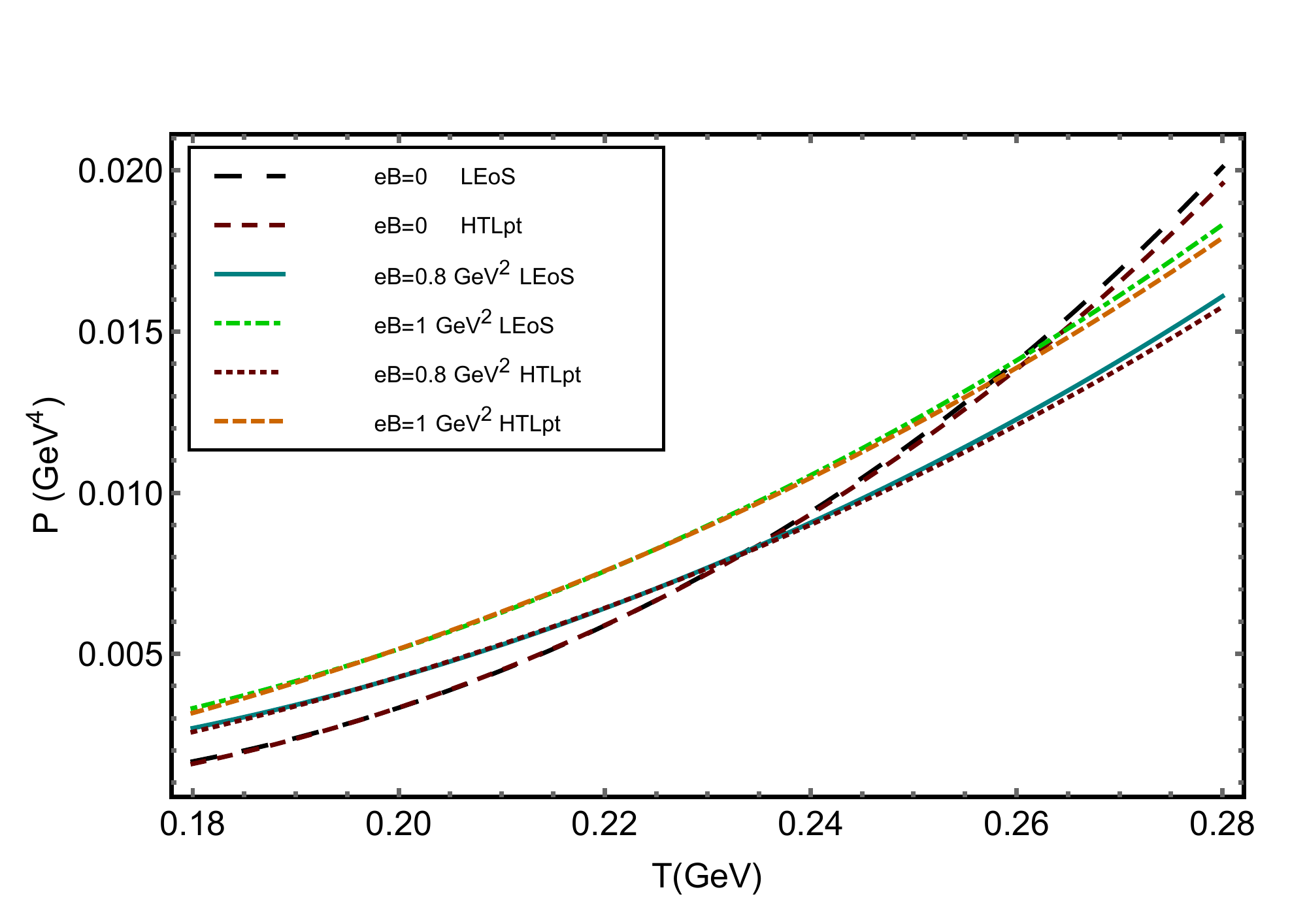}}
\hspace{4mm}
\subfloat{\includegraphics[height=6.50cm,width=5.50cm]{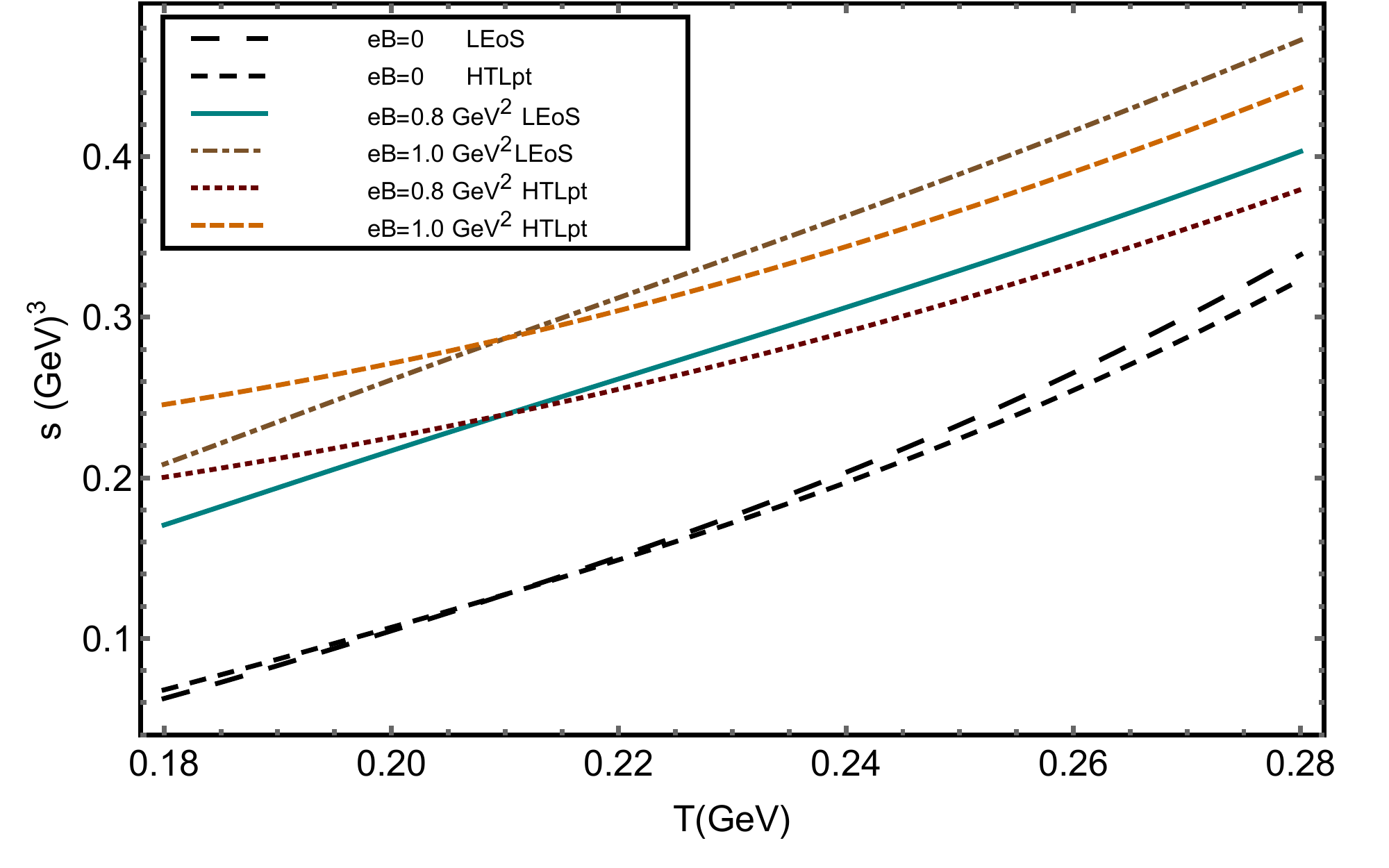}}
\caption{(color online) Temperature behavior of energy density (left panel), pressure (middle panel), and entropy density (right panel) for various values of magnetic fields are shown for two different EoSs.}
\label{101}
\end{figure*} 

Finally, the velocity of sound is another  fundamental quantity which is used to describe hot and dense QCD medium. The velocity of sound square , $c^{2}_{s}$,
\begin{equation}
c^{2}_{s}=\dfrac{\partial P}{\partial \varepsilon}\equiv
\dfrac{dP/dT}{d\varepsilon/dT}.
\end{equation}
Next, employing the expressions for the pressure and energy density, we obtain,
 \begin{align}
c_{s}^{2}&= \lbrace4\dfrac{T^{3}}{\pi^{2}}\nu_{g}PolyLog[4,z_{g}]\nonumber\\
&+\dfrac{T^{4}}{\pi^{2}}\nu_{g}PolyLog[3,z_{g}]\left(\partial_{T}\ln z_{g}\right) \nonumber\\
&-\mid q_{f}B\mid\dfrac{2T}{\pi^{2}}\dfrac{1}{3}\nu_qPolyLog[2,-z_{q}]\nonumber\\
&+\mid q_{f}B\mid\dfrac{2T}{\pi^{2}}\dfrac{1}{3}\nu_q\ln(1+z_{q})\left(\partial_{T}\ln z_{q}\right)\rbrace /\nonumber\\
&\lbrace\dfrac{12T^{3}}{\pi^{2}}\nu_{g}PolyLog[4,z_{g}]\nonumber\\
&+\left( T^{2}\partial_{T}\ln z_{g}\right)\dfrac{8T^{2}}{\pi^{2}}\nu_{g}PolyLog[3,z_{g}]\nonumber\\
&+T^{2}\left(\partial_{T}\ln z_{g}\right)^{2}\dfrac{T^{3}}{\pi^{2}}\nu_{g}PolyLog[2,z_{g}]\nonumber\\
&+T^{2}\left(\partial^{2}_{T}\ln z_{g}\right)\dfrac{T^{3}}{\pi^{2}}\nu_{g}PolyLog[3,z_{g}]\nonumber\\
&-\dfrac{4\mid q_{f}B\mid T}{\pi^{2}}\nu_qPolyLog[2,-z_{q}]\nonumber\\
&+5\mid q_{f}B\mid\left( T^{2}\partial_{T}\ln z_{q}\right)\dfrac{1}{\pi^{2}}\nu_q\ln(1+z_{q})\nonumber\\
&+\mid q_{f}B\mid T^{2}\left(\partial_{T}\ln z_{q}\right)^{2}\dfrac{T}{\pi^{2}}\nu_q\dfrac{z_{q}}{1+z_{q}}\nonumber\\
&+\mid q_{f}B\mid T^{2}\left( \partial^{2}_{T}\ln z_{q}\right)\dfrac{T}{\pi^{2}}\nu_q\ln(1+z_{q})\rbrace . 
\end{align}

\subsection{Discussions and comparison with other approaches}
\begin{figure*}
  \centering
  \hspace{-6.5mm}
\subfloat{\includegraphics[height=7.650cm,width=6.39cm]{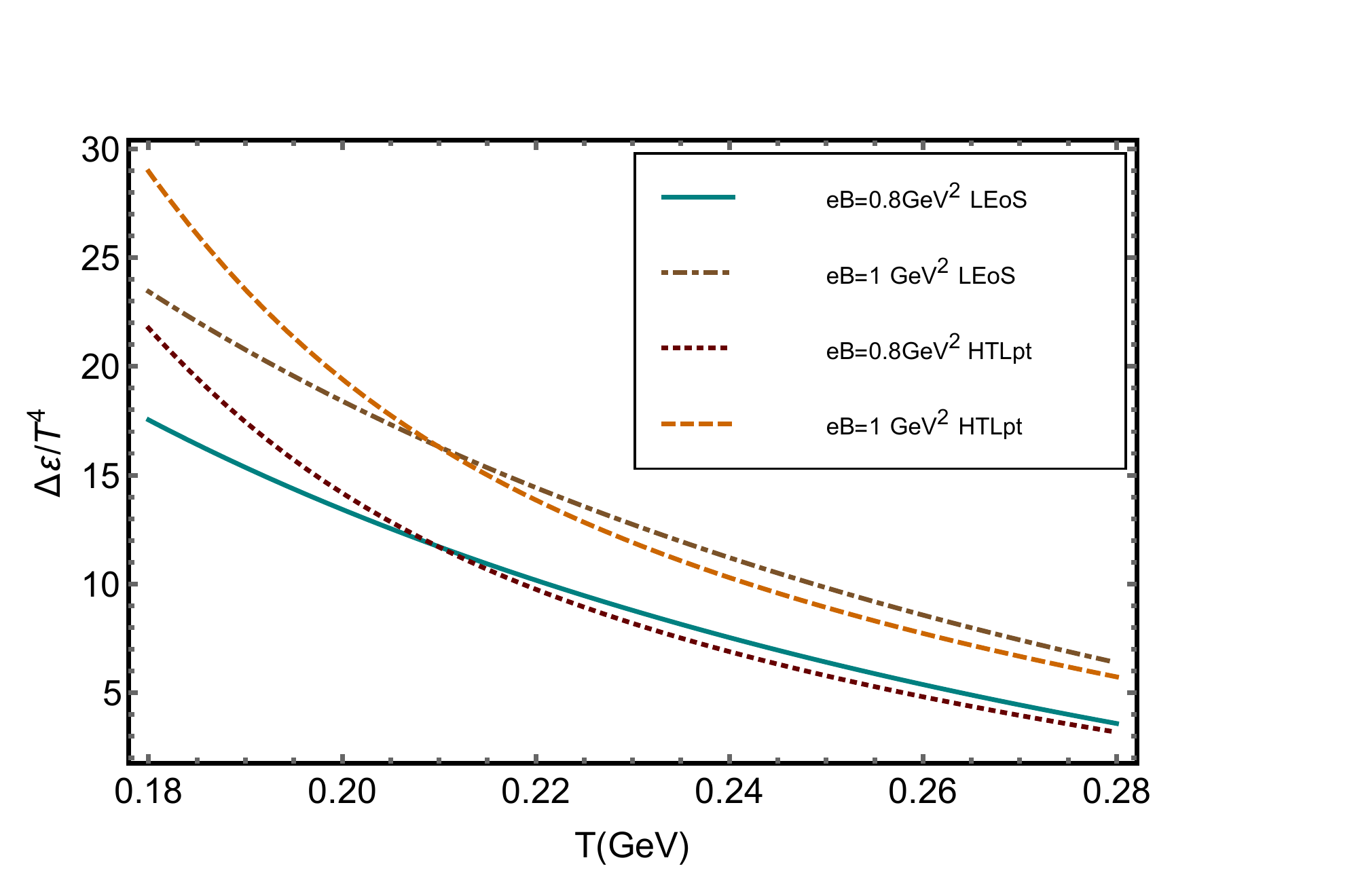}}
\hspace{-7mm}
\subfloat{\includegraphics[height=7.55cm,width=6.39cm]{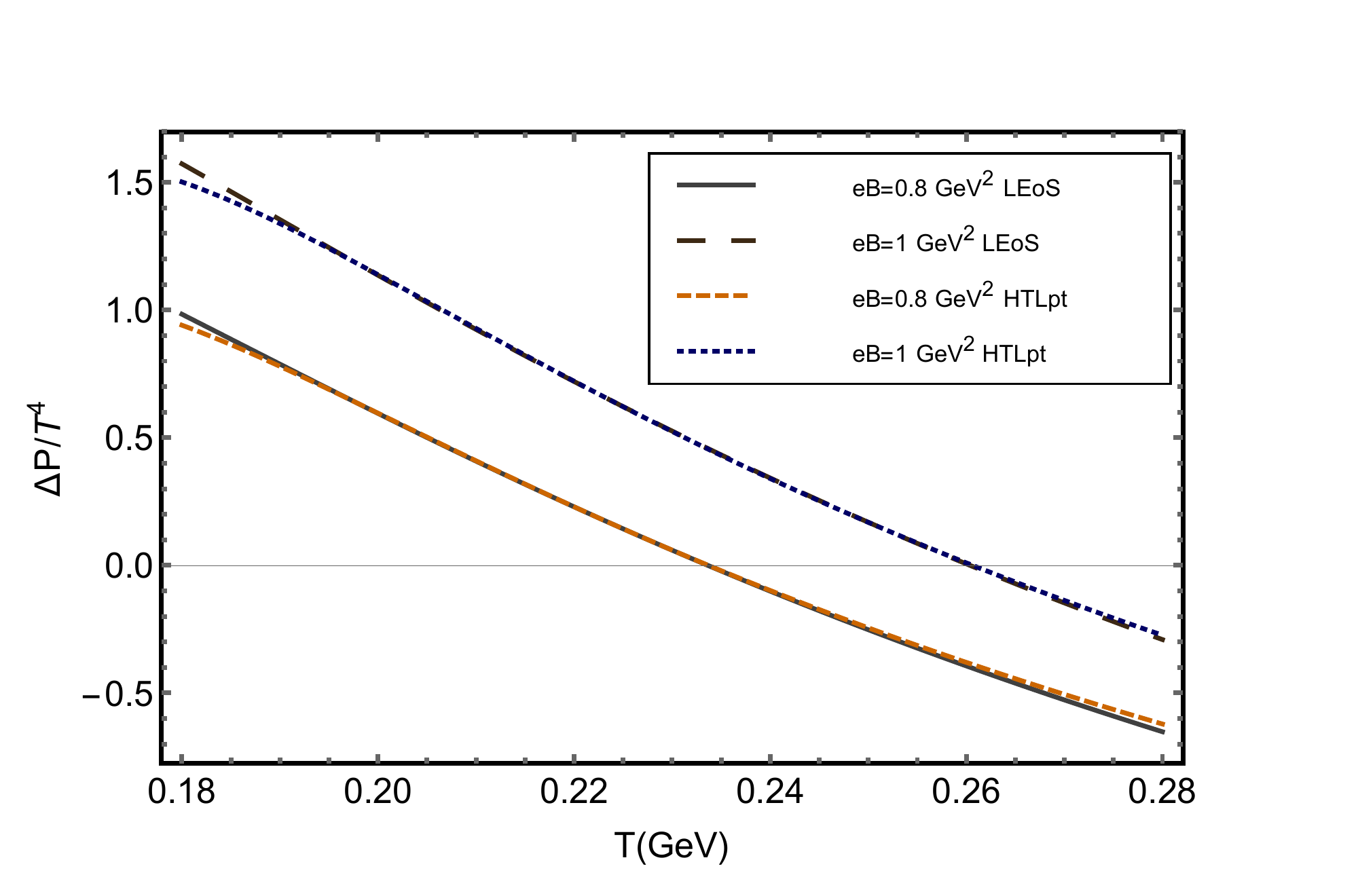}}
\hspace{-2mm}
\subfloat{\includegraphics[height=7.60cm,width=6.39cm]{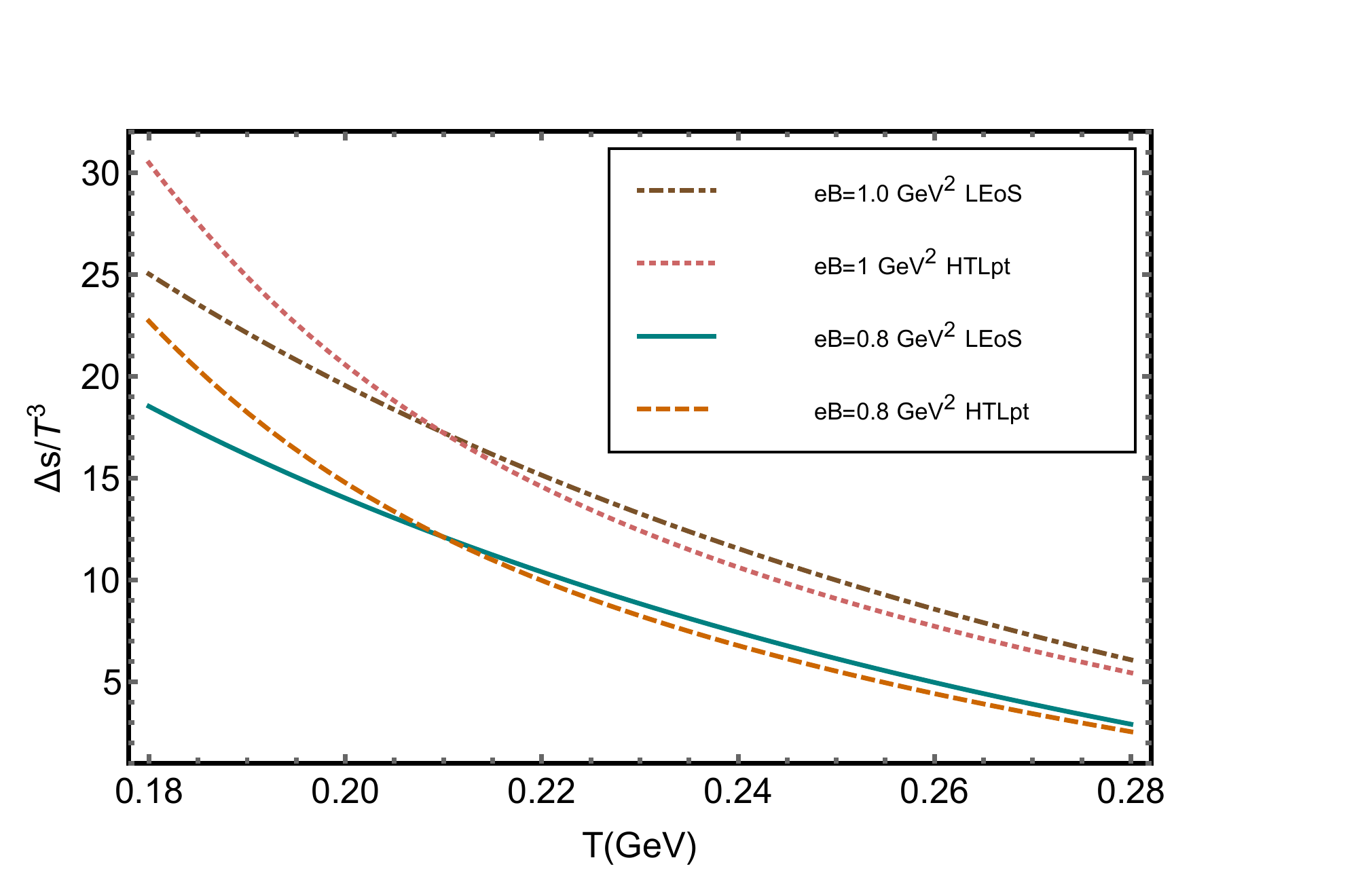}}
\caption{(color online) $\Delta\varepsilon/T^{4}$ (left panel), $\Delta P/T^{4}$ (middle panel), and $\Delta s/T^{3}$ (right panel) for various values of magnetic fields, plotted as a function of temperature. }
\label{102}
\end{figure*}

\begin{figure}[h]
\includegraphics[height= 7.9cm,width= 7.8cm]{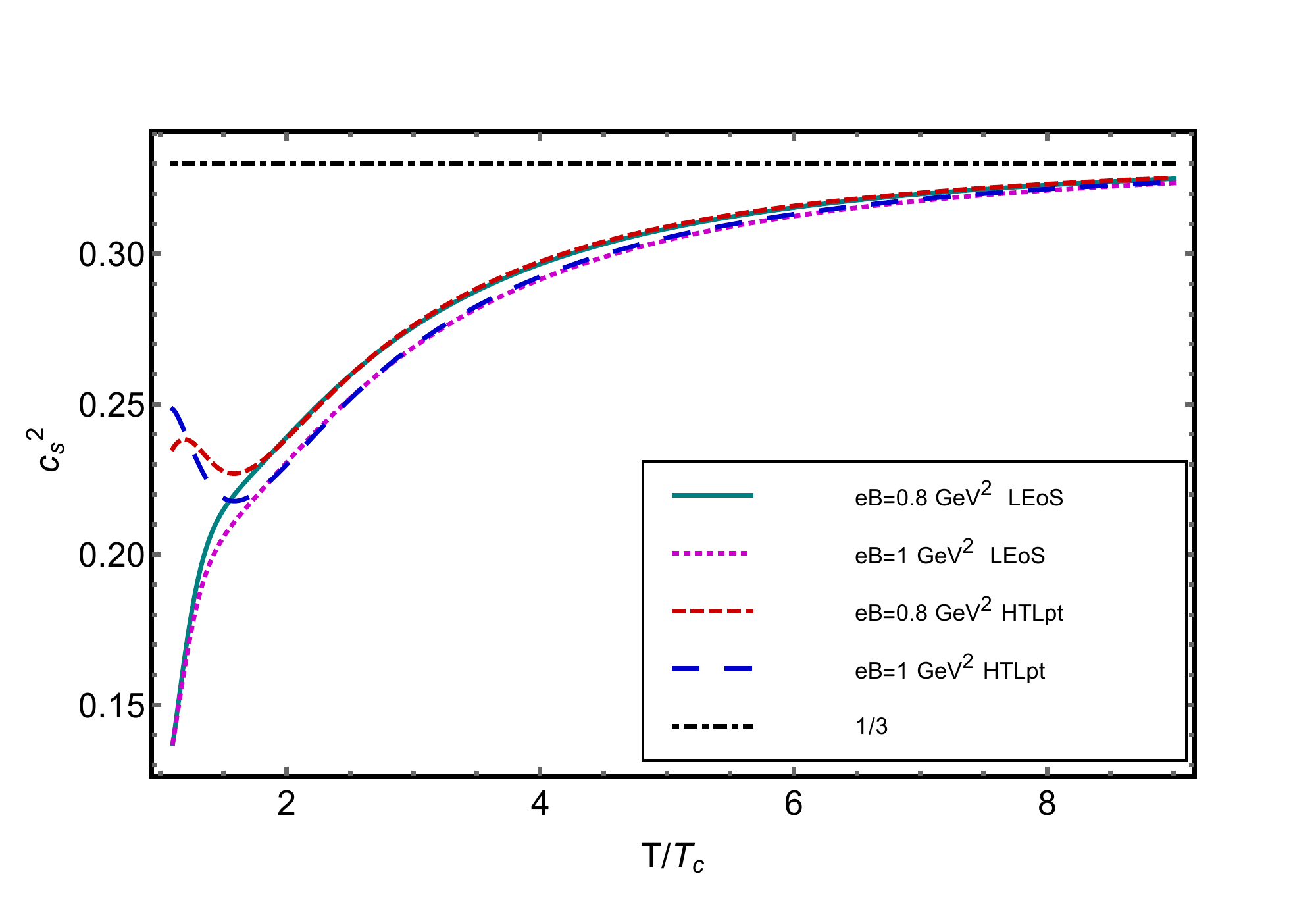}
\caption{(color online) Behavior of $c_{s}^{2}$ in strong magnetic field background for the two different EoSs. }
\label{103}
\end{figure}

\begin{figure}[h]
\includegraphics[height=7.80cm,width=8.80cm]{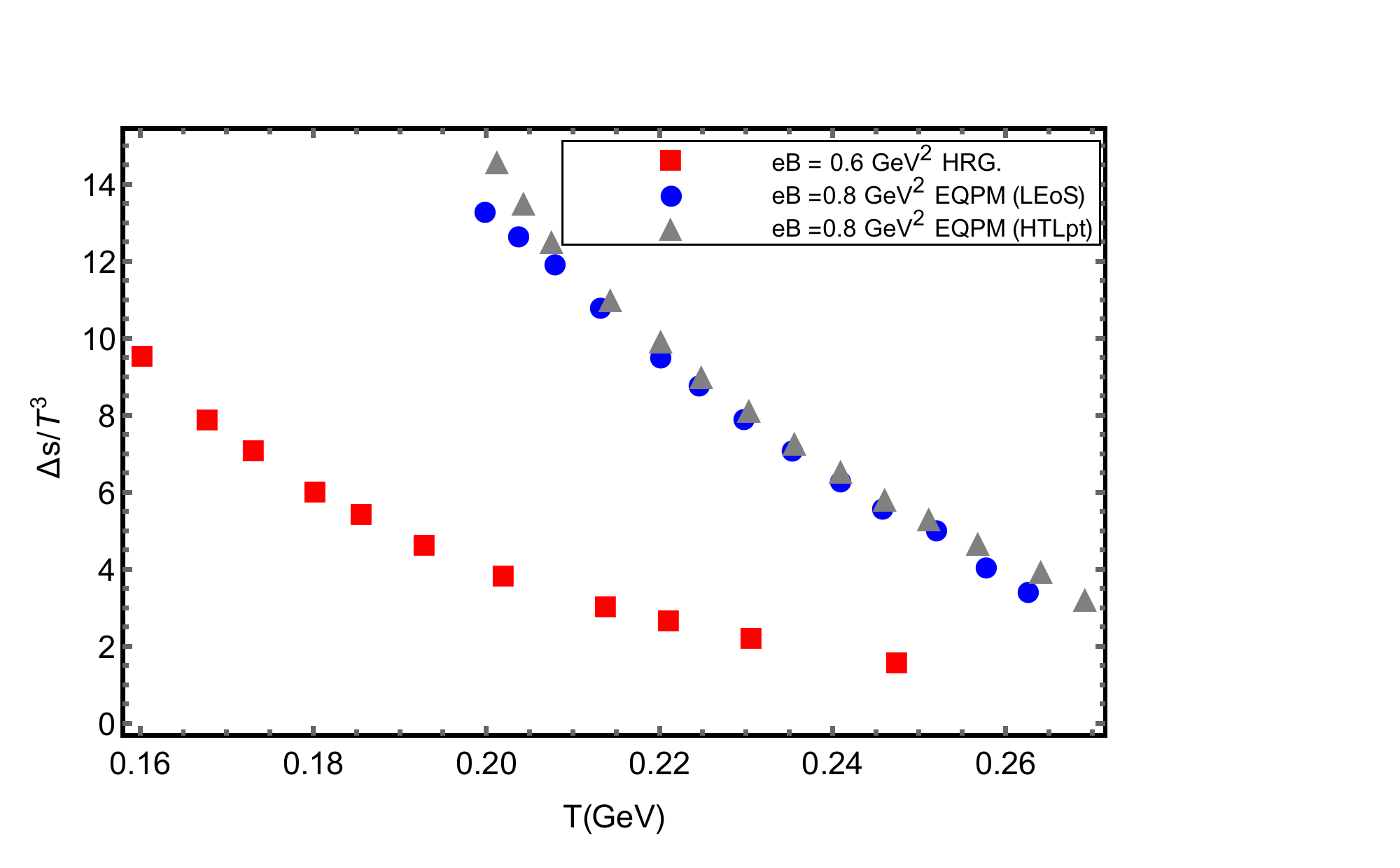}
\caption{Comparison of the behavior of $\Delta s/T^{3}$ as a function of temperature  for different magnetic fields. For $\mid eB\mid = 0.6$ GeV$^{2}$ the results of the HRG model~\cite{Bali:2014kia} is shown. }
\label{104}
\end{figure}

\begin{figure}[h]
  \subfloat{\includegraphics[height=8.20cm,width=8.5cm]{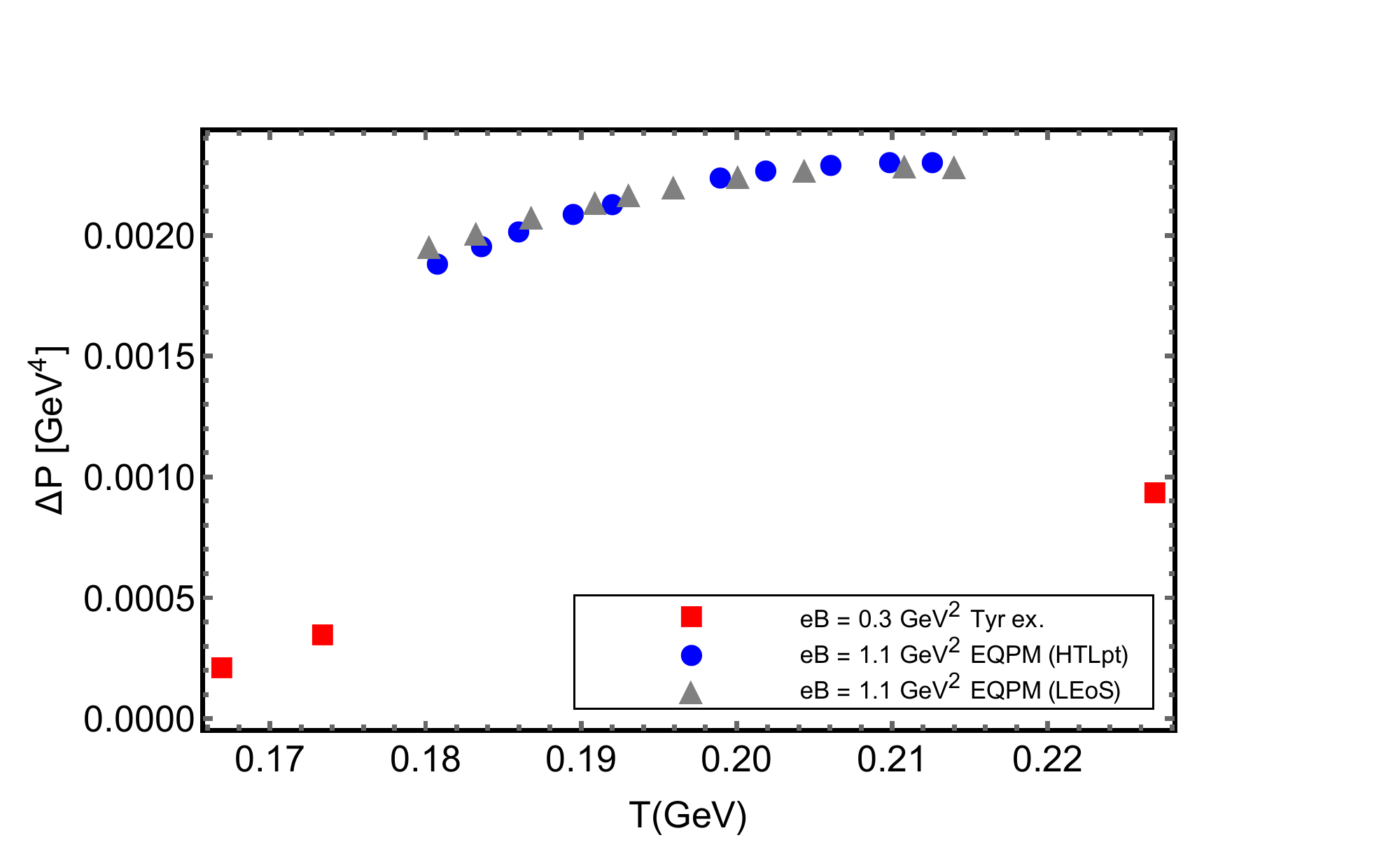}}
\caption{ Comparison of the behavior of change in pressure for two different values of magnetic fields. For $\mid eB\mid = 0.3$ GeV$^{2}$, the plot is obtained from the Taylor expansion of the pressure with respect to magnetic field~\cite{Levkova:2013qda}.}
\label{105}
\end{figure}

After obtaining the thermodynamic quantities, we have shown their explicit temperature dependence, with different EoSs
 and different values of the magnetic field in Figs.~\ref{101}-\ref{105}.  We observe that the values of pressure, energy density, and entropy density enhance with the increase in strength of the magnetic field.    
The increasing behavior of energy density ($\varepsilon$) with increasing temperature in different magnetic fields is shown in Fig.~\ref{101} (left panel) for both the EoSs. The observed behavior can be understood in the following way. Since in presence of magnetic field, $\varepsilon_{total}=\varepsilon+q \bf{M}\cdot\bf{B}$ where $q\bf{M}\cdot\bf{B}$ is the $\varepsilon_{field}$ where $\bf{M}$ is the magnetization.
 The temperature behavior of entropy density can be understood from Eq.~(\ref{19}) as  shown in the middle panel. The variation of the pressure as a function of temperature is shown in the right panel of Fig.~\ref{101}.
EoSs dependence of these quantities through the fugacity factor can also be distinguished from their respective  temperature behavior.  
 
The quantity $\Delta\varepsilon$, the difference between energy densities with and without the magnetic field, defines the increment of $\varepsilon$ in the presence of magnetic field. Temperature dependence of $\Delta\varepsilon/T^{4}$ is depicted in the Fig.~\ref{102} (left panel). 
We have also plotted the $\Delta P/T^{4}$ and $\Delta s/T^{3}$ as a function of temperature in the middle and right panel of the same figure. All the three  quantities  show decreasing pattern with increasing temperature. As expected, higher the magnetic field higher will be values of them.
 
In Fig.~\ref{103}, $c_{s}^{2}$ has been plotted as a function of the scaled temperature, $T/{T_{c}}$ for different magnetic field values.  The trend indicates that, $c_{s}^{2}$  will reach to its  Stefan-Boltzmann (SB) limit which is 1/3, only asymptotically.
For both EoS,   variation  of  $c_{s}^{2}$ with temperature is almost identical.
 At low temperature, the kink around $T_{c}$, shows that $c_{s}^{2}$ is sensitive to the particular choice of EoS, in that temperature range. Beyond the temperature range considered here and at higher temperatures  the higher Landau level corrections might play prominent role in understanding the temperature behavior of the thermodynamic quantities
  which is beyond the scope of present work and will be investigated in 
 the near future.

 We compare our results with those from HRG predictions. In~\cite{Bali:2014kia}, comparison of lattice results with HRG prediction for $\mid eB\mid = 0.6$ GeV$^{2}$ and other values of magnetic field were done. Fig.~\ref{104} shows the comparison of $\Delta s/T^{3}$ at $\mid eB\mid = 0.8$ GeV$^{2}$ from the extended EQPM 
 with existing result at $\mid eB\mid = 0.6$ GeV$^{2}$ from~\cite{Bali:2014kia}. The change in entropy density, decreasing trend with increasing temperature, obtained from the present calculation  agrees with the HRG predictions. The enhancement of entropy density with increasing magnetic field depicted in Fig.~\ref{101} justifies, the difference in the magnitude of the two plots.
         
 In~\cite{Levkova:2013qda}, the effects of an external magnetic field on the EoS of  the QGP are studied using numerical simulations of lattice QCD. We are comparing the change in pressure ($\Delta P$) from our model with this approach in Fig.~\ref{105}. We observe similar temperature behavior in the change in pressure as well. For $\mid eB\mid = 0.3$ GeV$^{2}$ the 
 increment in pressure due to the presence of magnetic field, appear to be compared to the present work at $1.1$ GeV$^{2}$. This fact can be explained from Eq.~(\ref{17}), which reveals the temperature dependence of increment in pressure, increases with higher values of magnetic field. The trend and quantitative behavior also follows Ref.~\cite{Bali:2014kia}, which
  have provided the same work with $\mid eB\mid = 0.2$ GeV$^{2}$ and $\mid eB\mid = 0.4$ GeV$^{2}$.
  
\section{The Debye mass and effective coupling in strong magnetic field}
Screening of color forces, in the hot QCD medium can be described in terms of Debye mass ($m_{D}$).  As we know plasma is the collection of both charged and neutral quasi-particles, which exhibit collective behavior. It's ability to shield out electric potential applied to it, can be measured in terms of Debye screening length (inverse of $m_{D}$). 
The conventional definition of Debye mass is given by the small momentum limit of the gluon self-energy~\cite{Rebhan, Shuryak}, which can also be realised within semi-classical transport theory~\cite{Mrowczynski, Yagi}. In the absence of magnetic field,  $m_{D}$ in terms of isotropic distribution function can be given as~\cite{Jamal:2017dqs}
\begin{equation}
m_{D}^{2}=-4\pi\alpha_{s}\int{\dfrac{d^{3}\vec{p}}{(2\pi)^{3}}\dfrac{df_{eq}}{dp}} ,
\end{equation}
with the equilibrium distribution function given by,
\begin{equation}
f_{eq}=2N_{c}f_{g}+ N_{f}(f_{q}+f_{\bar{q}}).
\end{equation}
Therefore In the absence of any magnetic field, we obtain
\begin{figure}[h]
  \subfloat{\includegraphics[height=7.5cm,width=8.2cm]{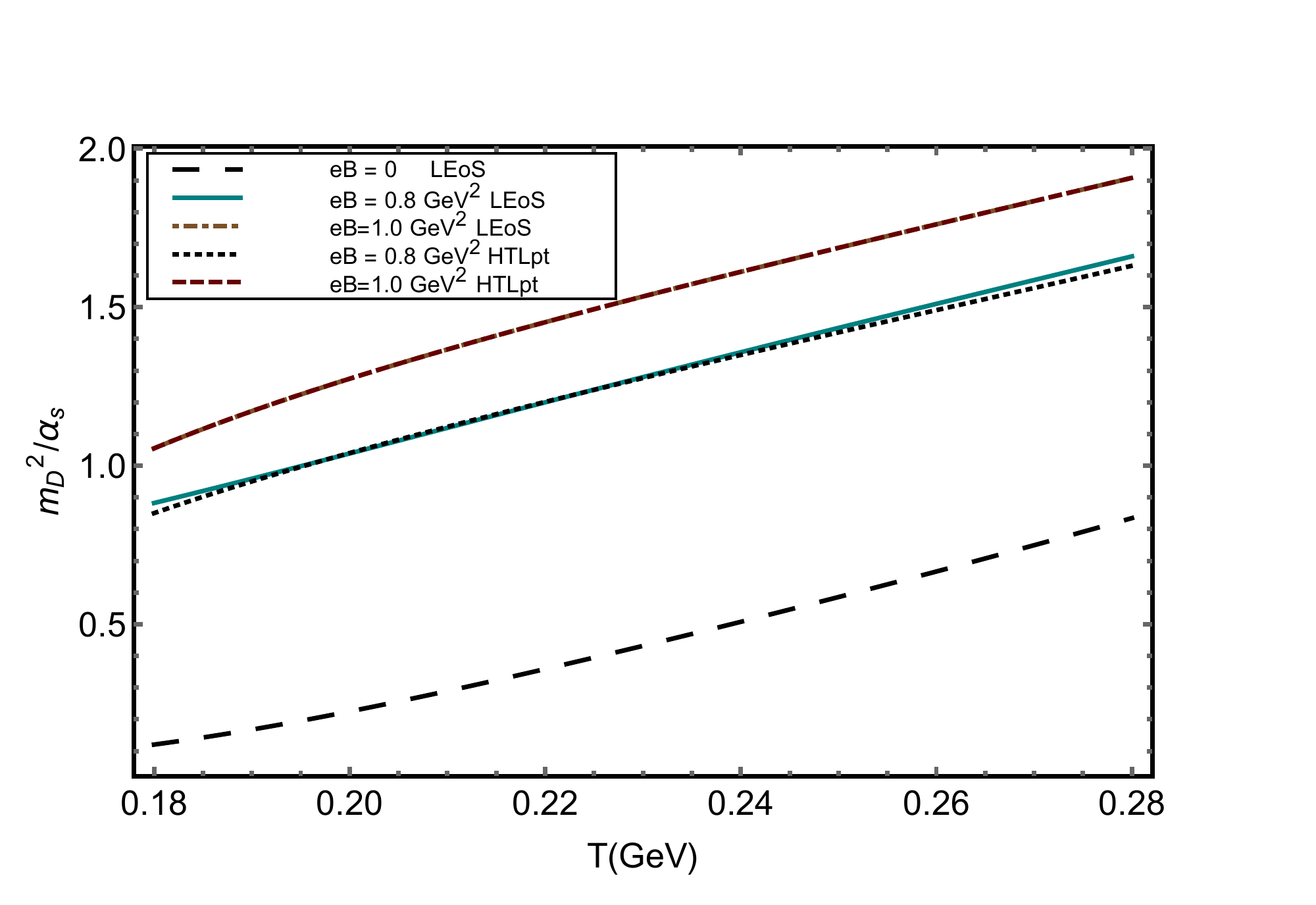}}
\caption{(color online) The ratio of Debye mass to the perturbative QCD running coupling constant as a function of temperature with and without magnetic fields.}
\label{106}
\end{figure}  
\begin{equation}
(m_{D}/\alpha_{s})=\dfrac{24T^{2}}{\pi}\left( PolyLog[2,z_{g}]-PolyLog[2,-z_{q}]\right), 
\end{equation}
$\alpha_{s}(T)$ is the running coupling constant at finite temperature taken from 2-loop QCD gauge coupling constants~\cite{Laine:2005ai}. 

Next, we are interested in computing the change in the Debye mass in presence of strong magnetic field.  This requires the definition of the Debye mass in the presence of the magnetic field in hot QCD. To that end, we can start with its definition in terms 
of the gluon self-energy:
\begin{equation}\label{26}
m_{D}^{2}=\Pi_{00}(\omega=0,\mid \vec{p}\mid \longrightarrow 0).
\end{equation}
The gluon self-energy gets modified in strong field background as~\cite{Bandyopadhyay:2016fyd},
\begin{equation}
\Pi_{00}(\omega=0,\mid \vec{p}\mid \longrightarrow 0)=\dfrac{g^{2}}{2\pi^{2} T}\mid eB\mid\int_{0}^{\infty}{dp_{z}f_{q}^{0}(1-f_{q}^{0})}.
\end{equation}
From this,  $m_{D}^{2}$ for quarks can be calculated as
 \begin{equation}
m_{D}^{2}=\dfrac{4\alpha_{s}}{\pi T}\mid eB\mid\int_{0}^{\infty}{dp_{z}f_{q}^{0}(1-f_{q}^{0})}.
\end{equation}
The gluonic contribution to the Debye mass will remain intact.
Also from kinetic theory approach, we can intuitively derive the expression for Debye screening mass in presence of magnetic field for perturbative QCD. Both of these approaches lead to the same expression for the Debye mass.

The intuitive realization of the screening for transport theory is as follows. In QED, charge density $\rho_{i}$ will induce due to the potential $V(x)$. From Maxwell equations,
\begin{align}
\bigtriangledown^{2}V(x)&=-\rho_{i}\nonumber\\
&= -m_{D}^{2}V(x).
\end{align}
Thus we can calculate the Debye mass in QED from the inhomogeneous Poisson equation. For non-abelian case the induced charge density for perturbative QCD~\cite{Jankowski:2015eoa} is 
\begin{equation}
\rho_{i}=2g\dfrac{eB}{2\pi}\dfrac{1}{2}\int_{-\infty}^{\infty}{\dfrac{dp_{z}}{(2\pi)}(f_{+}-f_{-})},
\end{equation}
where $f_{\pm}=\pm g\dfrac{df_{q\bar{q}}^{0}}{dp_{z}}V(x)$.

Following the same prescription as in QED, we will get Debye mass for of quark and anti-quarks as
\begin{equation}\label{31}
m_{D}^{2}=-4\alpha_{s}\dfrac{eB}{\pi}\int_{0}^{\infty}{dp_{z}\dfrac{df_{q}^{0}}{dp_{z}}}.
\end{equation} 
 
From Eqs.~(\ref{26}) and~(\ref{31}), we are getting the same expression for Debye mass for hot QCD system, in presence of strong magnetic background (with $N_{f}=3$ and $N_{c}=3$) as,  
\begin{equation}\label{32}
(m_{D}^{2}/\alpha_{s})=4\left( \dfrac{6T^{2}}{\pi}PolyLog[2,z_{g}]+\dfrac{3eB}{\pi}\dfrac{z_{q}}{1+z_{q}}\right) .
\end{equation} 
We have plotted the variation of the ratio of Debye mass to running coupling constant as a function of temperature in Fig.~\ref{106}. The temperature dependence of the ratio has an increasing trend with increasing temperature following from Eq.~(\ref{32}), which further enhances in the presence of a larger magnetic field. So we can infer, the effective dimensional reduction due to the strong magnetic field is contributing a significant effect on the screening to coupling constant ratio.

Next, we can define the effective coupling in presence of strong magnetic field background. For ideal EoS ($z_{q,g}=1$), representing the ultra-relativistic non-interacting quarks and gluons, we can rewrite the definition of Debye mass as,
\begin{equation}\label{33}
(m^{2}_{D})_{Ideal}=\alpha_{s}(T) 4\pi\left( T^{2}+\dfrac{3eB}{2\pi^{2}}\right). 
\end{equation}

From Eqs.~(\ref{32}) and~(\ref{33}), we can define the effective running coupling $\alpha_{eff}(T)$, so that $m^{2}_{D}=\alpha_{eff}(T,z_{q},z_{g}) 4\pi\left(T^{2}+\dfrac{3eB}{2\pi^{2}}\right) $ and can be expressed as,
\begin{equation}\label{34}
\alpha_{eff}=\alpha_{s}(T)\dfrac{\left( \dfrac{6T^{2}}{\pi^{2}}PolyLog[2,z_{g}]+\dfrac{3eB}{\pi^{2}}\dfrac{z_{q}}{(1+z_{q})}\right)}{\left( T^{2}+\dfrac{3eB}{2\pi^{2}}\right)}.
\end{equation}
\begin{figure}[h]
  \subfloat{\includegraphics[height=7.50cm,width=8.2cm]{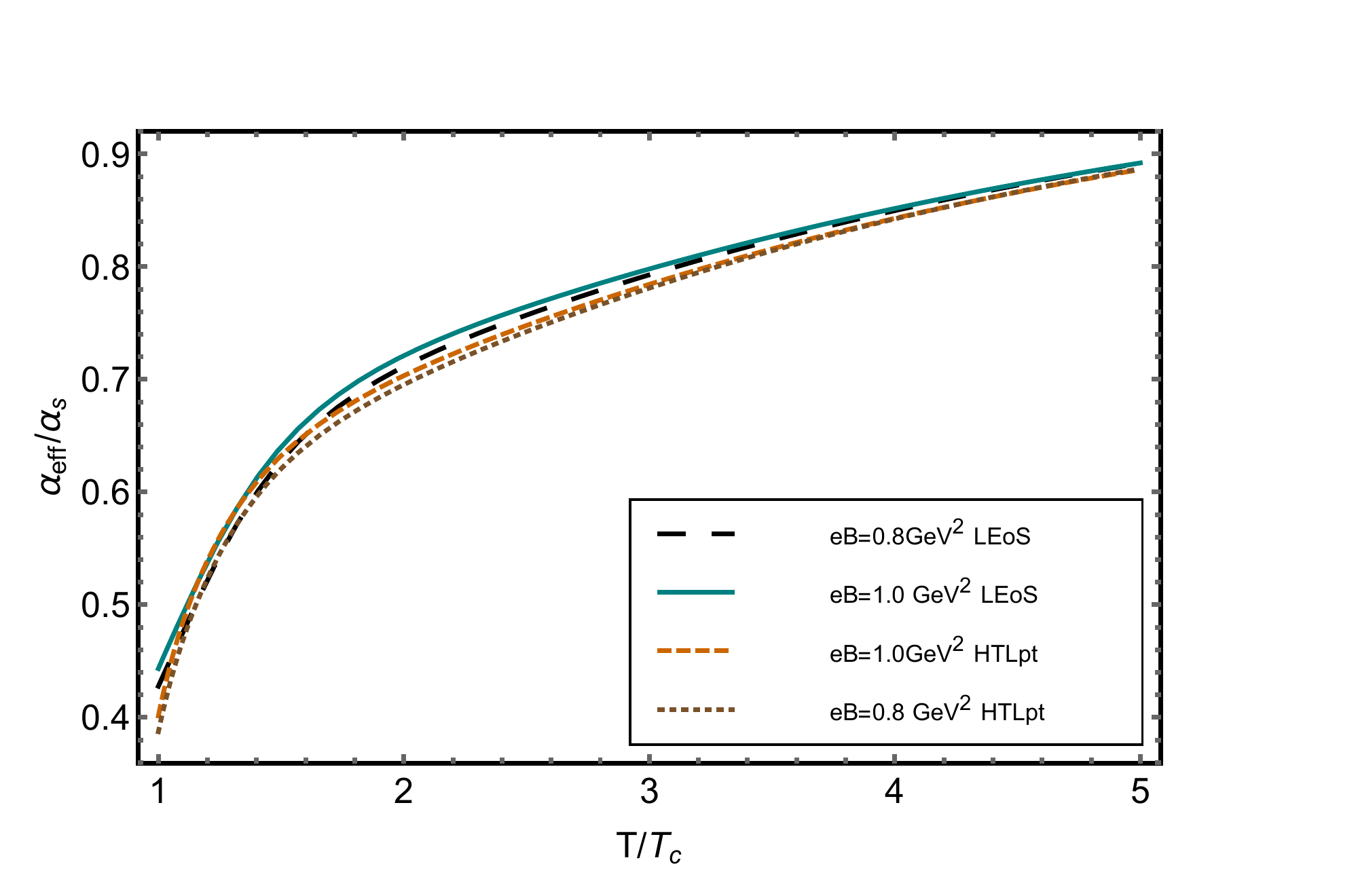}}
\caption{(color online) Effective coupling constant in presence of the  magnetic field in LLL  approximation}.
\label{107}
\end{figure}

The temperature behavior of the ratio of $\alpha_{eff}/\alpha_{s}$, considering only the lowest Landau states in presence of magnetic field is shown in Fig.~\ref{107}. At lower temperature, $\alpha_{eff}$ is much smaller than $\alpha_{s}$ due to the screening effect provided by the strongly interacting quasi-partons under the EQPM scheme. Asymptotically, the ratio is approaching unity since at high temperature the quasi-partons will start behaving like free particles. The ratio $\alpha_{eff}/\alpha_{s}$ is showing a small but quantitative change in the temperature dependence due to increasing magnetic field. At very high temperature, higher order Landau corrections are needed to be involved. The temperature range in which lowest Landau levels are dominant can fix from the LLL approximation. Effective coupling has significant importance in the extended EQPM description of transport properties in QGP. Being an essential dynamical input for transport processes, the effective coupling controls the behavior of transport parameters critically. We will use this concept in the calculation of longitudinal electrical conductivity in presence of the magnetic field in the next section.

\section{Longitudinal conductivity of QCD at high temperature in strong field background }
From the EQPM in the strong magnetic field background ($B=B\hat{z}$), we investigate the longitudinal conductivity of the QCD at high temperature. We are working in the regime $\alpha_{s}eB\ll T^{2}\ll eB$ where $\alpha_{s}$ is the running coupling constant of QCD. The inequality means that the regime under consideration is weakly coupled and the magnetic field background is considerably strong, that allows the LLL approximation. To get an estimate of the transport coefficients in kinetic theory approach, we need to start from relativistic transport equation, which quantifies the rate of change of distribution function in terms of collision integral. This collision term includes all the binary elastic process among quarks/anti-quarks and gluons along with the annihilation and pair production mechanisms. The latter is dominant in the presence of magnetic field since their rate is proportional to $\alpha_{s}$ whereas that of the binary process is proportional to $\alpha_{s}^{2}$. The prime focus of this work is on the dominant 1 $\rightarrow$ 2 scattering (gluon to quark/anti-quark pair). We are considering quasi-partons distribution function from the extended EQPM in strong field background from Eq.~(\ref{7}) in order to take the interacting medium into account. The longitudinal current density in the applied electric field  $\vec{E}=E\hat{z}$ is obtained as, 

\begin{equation}\label{35}
J_{z}=2eD(R)\dfrac{eB}{(2\pi)}\int{\dfrac{dp_{z}}{(2\pi)}v\delta f_{q}},
\end{equation}
where 2 factor comes from the identical behavior of quark and anti-quark. $v$ is the longitudinal velocity  in the direction of $\bf{E}$, $v=\frac{p_{z}}{E_{p}}$ and $D(R)$ is the color representation of the quarks. Calculation of longitudinal conductivity from Eq.~(\ref{35}) is straightforward. The quantity $\delta f_{q}$ is the change from the local momentum distribution function of quasi-quarks in the following way, 
\begin{equation}\label{36}
f_{q}(p_{z})= f_{q}^{0}+\delta f_{q},
\end{equation}
with 
\begin{align}
\delta f_{q}(p_{z})&= \beta f_{q}^{0}\dfrac{z^{-1}\exp{\beta E_{p_{z}}}}{(1+z^{-1}\exp{\beta E_{p_{z}}})}\chi(p_{z})\nonumber\\ \nonumber\\
&= \beta f_{q}^{0}(E_{p_{z}})(1-f_{q}^{0})\chi(p_{z}).\label{37}
\end{align}
where $f_{q}^{0}$ is defined in Eq.~(\ref{7}). Here, $\chi(p_{z})$ is the response function in presence of applied electric field. Since the response of quark and antiquark in electric field is exactly opposite, $\chi(p_{z})$ is an odd function of $p_{z}$, that is $\chi(-p_{z})=-\chi(p_{z})$. Quasi-quark distribution function dynamics can be described from the Boltzmann equation~\cite{Heinz:1984yq} as  
\begin{equation}\label{38}
\dfrac{\partial f_{q}}{\partial t}+eE\dfrac{\partial f_{q}}{\partial p_{z}}=C(f_{q}).
\end{equation} 
Here $C(f_{q})$ represents the collision integral and for leading 1 $\rightarrow$ 2 process $( k\longrightarrow p+p^{'} )$ this have the following form~\cite{Hattori:2016lqx},

\begin{align}\label{39}
C(f_{q})=&8\pi^{2}\alpha_{eff}C_{2}\int^{\infty}_{-\infty}{\dfrac{dp^{'}_{z}}{2E_{p}2E_{p^{'}}}}\int{\dfrac{d^{2}k_{\perp}}
{(2\pi)^{2}2E_{k}}}\dfrac{k_{\perp}^{2}}{\mid k\mid^{2}}\nonumber\\
&\times\left( E_{p}E_{p^{'}}+p_{z}p_{z^{'}}+m^{2}\right)\delta(E_{k}-E_{p}-E_{p^{'}})\nonumber\\
&\times f_{q}{0}(E_{p})f_{q}^{0}(E_{p^{'}})\left( 1+f_{g}^{0}(E_{k})\right)\nonumber\\
&\times\beta\exp{\left( -\dfrac{k^{2}_{\perp}}{2\mid eB\mid}\right) }\left( \chi_{q}(p_{z}^{'})-\chi_{q}(p_{z})\right),
\end{align}
with  $k_{z}=p_{z}+p_{z^{'}}$. Here $C_{2}$ is the Casimir factor and $\alpha_{eff}$ is the effective coupling constant, which incorporates the effects of EoS as given in Eq.~(\ref{34}). This expression is obtained using the equation of detailed balance of quasi-parton distribution functions in the basic definition of collision integral. Details of the calculations are shown in the Appendix. 
Solving the delta function by expanding each terms and using the condition $k_{z}=p_{z}+p_{z^{'}}$ we end up with  
\begin{align}\label{40}
\delta( E_{k}-E_{p}&-E_{p^{'}})= 2 ( E_{p}+E_{p^{'}} )\nonumber\\
&\times\delta\left(k_{\bot}^{2}-( E_{p}+E_{p^{'}})^{2}+(p_{z}+p_{z^{'}})^{2}\right) .
\end{align}

Exponential factor in the Eq.~(\ref{39}) tends to 1 because of the influence of very strong magnetic field. Using Eq.~(\ref{40}) we can do the $k^{2}_{\bot}$ integration substituting $E_{k}=E_{p}+E_{p^{'}}$, from $\delta$-function properties. Finally the collision integral becomes,

\begin{align}\label{41}
C(f_{q})&=\alpha_{eff}C_{2}\int_{-\infty}^{\infty}{\dfrac{dp^{'}_{z}}{E_{p}E_{p^{'}}}}( E_{p}E_{p^{'}}+p_{z}p_{z^{'}}+m^{2})\beta\nonumber\\
&\times\left( \dfrac{( E_{p}+E_{p^{'}})^{2}-(p_{z}+p_{z^{'}})^{2}}  {(E_{p}+E_{p^{'}})^{2}}\right) f_{q}^{0}(E_{p^{'}})\nonumber\\
&\times f_{q}^{0}(E_{p})( 1+f_{g}^{0}(E_{k}))\left( \chi_{q}(p_{z}^{'})-\chi_{q}(p_{z})\right)  . 
\end{align}
We can use this collision integral in Boltzmann equation in the homogeneous uniform electric field ($\dot{p}_{z}=eE$) (Eq.~(\ref{38})) to calculate the response function ($\chi(p_{z})$). Longitudinal conductivity can be then derived from this quantity conveniently, using Eq.~(\ref{35}). Here, we consider the Boltzmann equation at very near equilibrium $(\partial_{t}f_{q}=0)$. Under this approximation, the additional part of the energy dispersion ($T^{2}\partial_{T} \ln(z_{g/q})$) will not enter in the analysis through space-time derivative. This fact will also be respected by the collision term. The medium effects will only enter through the distribution function and the effective hot QCD coupling constant.\\

   Being motivated by the recent work Ref.~\cite{Hattori:2016cnt}, we are interested in calculating the longitudinal conductivity, in which the dominant contribution comes from the quarks of the momentum of order $T$. So, we are focused in the $p_{z^{'}}\sim 0$ regime. In this regime integrand of $\chi(p_{z^{'}})$ (Eq.~(\ref{41})) is an even function. Recall that $\chi(p_{z^{'}})$ is an odd function, which results in vanishing integral with $\chi(p_{z^{'}})$. From above assumptions $\chi(p_{z})$ can solve from Eqs.~(\ref{38}) and~(\ref{41}) and has the following form
\begin{equation}\label{42}
\chi(p_{z})=\dfrac{eEp_{z}(1-f^{0}_{q}(E_{p}))}{E_{p}C_{2}\alpha_{eff}I},
\end{equation}
where,
\begin{align}\label{43}
I&=\int_{-\infty}^{\infty}{dp_{z^{'}}\dfrac{f^{0}_{q}(E_{p^{'}})}{E_{p^{'}}E_{p}}(1+f^{0}_{q}(E_{p}+E_{p^{'}}))}\nonumber\\
&\times\dfrac{\left( (E_{p}+E_{p^{'}})^{2}-(p_{z}+p_{z^{'}})^{2}\right) (E_{p}E_{p^{'}}+p_{z}p_{z^{'}}+m^{2})}{(E_{p}+E_{p^{'}})^{2}}.
\end{align}
 The longitudinal conductivity ($\sigma^{L}_{eff}$) can be obtained from Eqs.~(\ref{35}),~(\ref{37}),~(\ref{42}) and given by
\begin{equation}\label{44}
\sigma^{L}_{eff}=(\dfrac{eB}{2\pi})\dfrac{2e^{2}D(R)}{C_{2}\alpha_{eff}}\int_{-\infty}^{\infty}{\dfrac{dp_{z}}{(2\pi)}\dfrac{p_{z}^{2}\beta f^{0}_{q}(1-f_{q}^{0})^{2}}{E_{p}^{2}I}}.
\end{equation}
In $p_{z^{'}}\sim 0$ regime, considering small $m$ limit we have, 
\begin{align}
&\dfrac{\left( (E_{p}+E_{p^{'}})^{2}-(p_{z}+p_{z^{'}})^{2}\right) (E_{p}E_{p^{'}}+p_{z}p_{z^{'}}+m^{2})}{(E_{p}+E_{p^{'}})^{2}}\nonumber\\
&\simeq \dfrac{\left( (E_{p}^{2}+2E_{p}m+m^{2}-p^{2}_{z})\right)(E_{p}m+m^{2})}{(E_{p}+m)^{2}}\nonumber\\
&= 2 m^{2}. 
\end{align}
 The above approximation helps us to solve Eq.~(\ref{43}), analytically. Following this we have
\begin{equation}
I=\dfrac{4}{E_{p}}\int_{0}^{\infty}{\dfrac{dp_{z^{'}}}{E_{p^{'}}}\dfrac{z_{q}}{1+z_{q}}(1+f^{0}_{g}(E_{p}))}m^{2},
\end{equation}
where through $z_{g}$ and $z_{q}$ the EoS effects are entering in the calculation. Asymptotically $z_{g,q}$ approaches unity. Finally, performing the integration we end up with
\begin{align}\label{47}
I= \dfrac{4}{E_{p}}m^{2}\dfrac{z_{q}}{1+z_{q}}\left( 1+f^{0}_{g}(E_{p})\right) \ln(T/m).
\end{align}
Putting Eq.~(\ref{47}) into Eq.~(\ref{44}), we can obtain the expression for conductivity as, 
\begin{align}
\sigma_{eff}^{L}= &\dfrac{eB}{2\pi}\dfrac{2e^{2}D(R)}{C_{2}\alpha_{eff}}\int_{-\infty}^{\infty}{\dfrac{dp_{z}}{2\pi}\lbrace\dfrac{ e^{\beta p}z_{q}^{-2}(z_{g}^{-1}e^{\beta p}-1)}{z_{g}^{-1}(z^{-1}_{q}e^{\beta p}+1)^{3}}}\nonumber\\
&\times\dfrac{p_{z}^{2}\beta}{E_{p}}\dfrac{z_{q}+1}{z_{q}}\dfrac{1}{4 m^{2}(1+f^{0}_{g}(E_{p}))\ln(T/m)}\rbrace.
\end{align}

 Since we are interested in temperature range $T > T_{c}$ fugacity is always greater than zero, but less than 1. Applying this fact and performing the integration, longitudinal conductivity at high temperature becomes
\begin{align}\label{49}
\sigma_{eff}^{L}=&\dfrac{eB}{2\pi}\dfrac{e^{2}D(R)}{\pi C_{2}\alpha_{eff}} \dfrac{T}{\ln (T/m)}\dfrac{(z_{q}+1)}{2z_{q}}\nonumber\\
&\times\dfrac{1}{4}\lbrace \dfrac{(z_{q}+z_{g})-(z_{g}-z_{q})\ln(z_{q})}{z_{q} m^{2}} \rbrace.
\end{align}
The impact of collective excitation of quasi-partons in the longitudinal electrical conductivity is embedded in Eq.~(\ref{49}) through the $z_{g},z_{q}$ and the effective coupling constant $(\alpha_{eff})$. Substituting $\alpha_{eff}$ from Eq.~(\ref{34}), the longitudinal electrical conductivity in presence of strong magnetic field, from extended EQPM can be expressed as
\begin{align}
\dfrac{\sigma^{L}_{eff}}{T}=&\dfrac{eB}{2\pi}\dfrac{e^{2}D(R)}{\pi C_{2}\alpha_{s}}\dfrac{(z_{q}+1)( T^{2}+\dfrac{3eB}{2\pi^{2}}) }{ \left( \dfrac{6T^{2}}{\pi^{2}}PolyLog[2,z_{g}]+\dfrac{3eBz_{q}}{\pi^{2}(1+z_{q})}\right) }\nonumber\\
&\times\dfrac{1}{8 m^{2}}\left( \dfrac{(z_{q}+z_{g})-(z_{g}-z_{q})\ln(z_{q})}{z_{q}^{2}\ln (T/m)}\right).
\end{align}
The ideal situation $(z_{q/g}=1)$, with quarks/anti-quarks and gluons without the hot QCD medium effects, conductivity simply becomes 
\begin{equation}
\dfrac{\sigma^{L}_{Ideal}}{T}=\dfrac{eB}{2\pi}\dfrac{e^{2}D(R)}{2\pi C_{2}\alpha_{s}}\dfrac{1}{m^{2}\ln (T/m)}.
\end{equation}
\begin{figure}[t]
 \subfloat{\includegraphics[height=7.5cm,width=8.2cm]{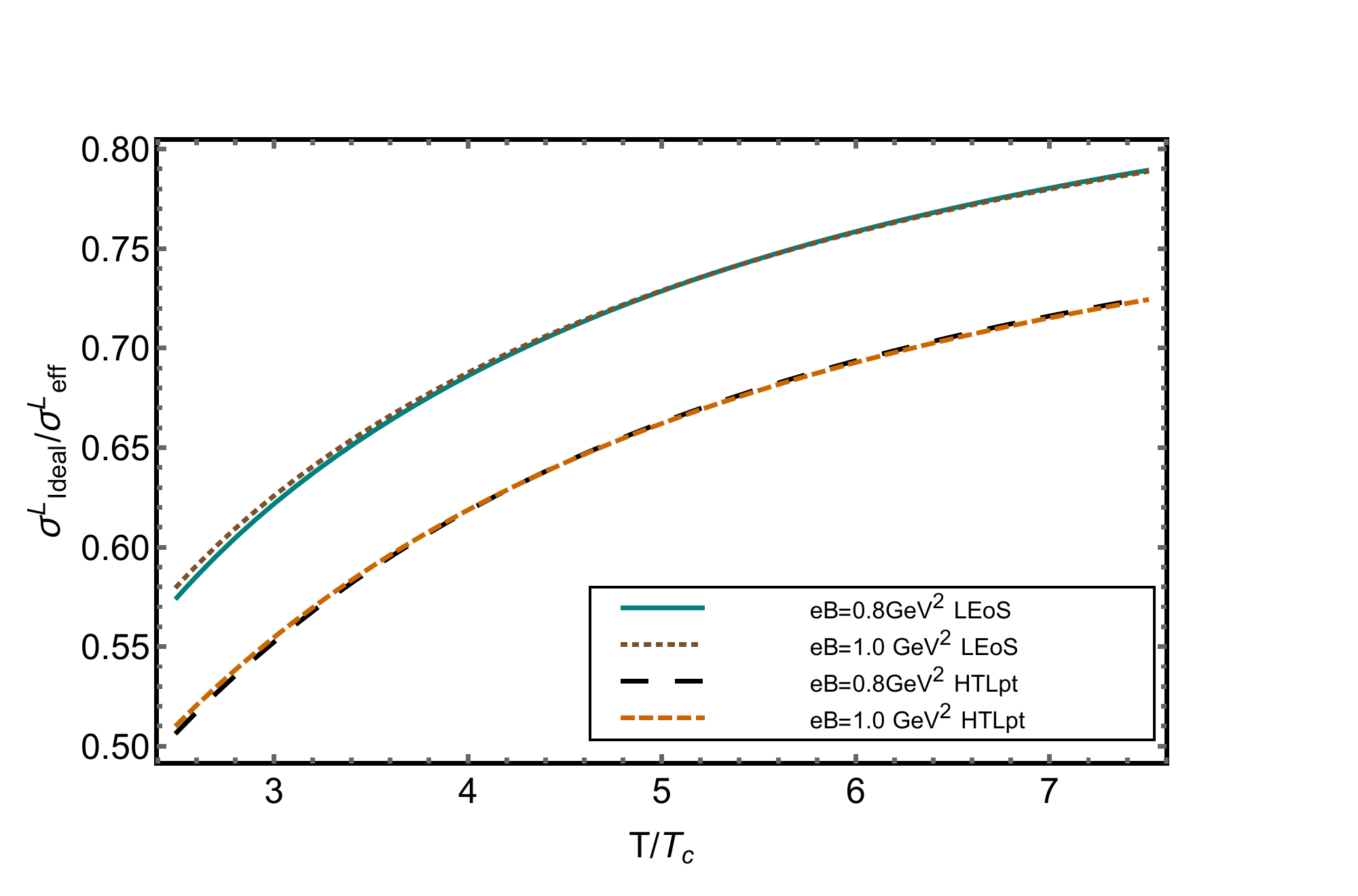}}
 \caption{(color online) Dependence of EoS on the longitudinal conductivity in strong magnetic field.}
\label{108}
\end{figure}
We plotted the variation of $\sigma^{L}_{Ideal}/\sigma^{L}_{eff}$ with $T/T_{c}$. Hot QCD medium effects in the longitudinal electrical conductivity are shown in Fig.~\ref{108}. The medium effects are entering through quasi-parton distribution functions along with the effective coupling. Since at lower temperature $\alpha_{eff}$ is lower than  $\alpha_{s}$ (Fig.~\ref{107}), we have larger value of $\sigma^{L}_{eff}$ with respect to $\sigma^{L}_{Ideal}$. The ratio ($\sigma^{L}_{Ideal}/\sigma^{L}_{eff}$) approach to unity as $z_{q/g}\rightarrow 1$. The EoSs dependence can also be seen from the Fig.~\ref{108}. Again for very high temperature, higher order corrections are needed to be included. Considering the magnitude of magnetic field, one can fix the temperature range in which lowest Landau states dominate, from LLL approximation. 

\section{Conclusion and Outlook}
In conclusion, we have extended the effective quasi-particle model (EQPM) for hot QCD in strong magnetic field. 
The impact of the magnetic field is included through fermionic quantization in the form of Landau levels. This modifies not only the momentum distributions of the quasi-quarks and anti-quarks but also their energy dispersions.
Consequently, the hot QCD thermodynamics gets significant modifications in strong field limit. The thermodynamic quantities such as the pressure, the energy density, the entropy density, and the speed of sound, for the hot QCD medium, have been computed, employing the extended EQPM and compared them against the estimates from other approaches. 
The energy density, the pressure, and the entropy density have shown visible increment in the presence of the magnetic field in the temperature range 180 MeV- 280 MeV as compared to the case where the field is absent. 
As expected, the square of the speed of sound has seen to approach in Stefan-Boltzmann limit only asymptotically.

Further, the Debye mass and consequently the hot QCD coupling constant in the presence of magnetic field has been obtained by adopting the definition of the Debye mass obtained from gluon self-energy. The Debye mass is found to be quite sensitive to the magnetic field. Finally, as an implication of the extended EQPM, the leading order term in the longitudinal electrical conductivity of hot perturbative QCD medium has been estimated while focusing on the quark dominating regime of the conductivity. The EoS (hot QCD medium) dependence on the longitudinal conductivity is seen to be very significant in the chosen range of the temperature which was entering through the effective coupling and the quasi-parton distribution functions. Finally, various predictions of the present work, either on the thermodynamic quantities or the effective coupling turned out to be consistent with other parallel approaches. The electrical conductivity (longitudinal) is seen to be sensitive to both the EoS and the magnetic field.
 
Throughout the calculation, the LLL approximation ($\mid eB\mid\gg T^{2}$) has been considered. For higher temperatures (for a fixed magnetic field) higher order Landau level corrections are non-negligible and can play a prominent role in understanding the transport properties of the hot QCD medium. This will be a matter of future investigations. In addition, we intend to estimate the  other transport coefficients such as shear and bulk viscosities in the strong magnetic field background with the extended EQPM in the near future along with investigating the Hall conductivity for 1 $\rightarrow$ 2 and 2 $\rightarrow$ 2 process in the QGP/hot QCD medium in the strong magnetic field.
 
\section*{acknowledgments}
V.C. would like to acknowledge SERB, Govt. of India for the Early
Career Research Award (ECRA/2016). We are thankful to Sukanya Mitra and M.Yousuf Jamal for helpful discussions and suggestions. We are indebted to the people of India for generous support for the research in basic sciences.

\appendix
\section{ Calculation of collision integral with quasiquark distribution function}
In the strong magnetic field background, 1 $\rightarrow$ 2 processes are kinematically possible. From the basic definition of collision integral, for the process~\cite{Hattori:2016lqx}
\begin{align}
C(f_{q})=&\int{\dfrac{d^{2}p^{'}_{z}}{(2\pi)^{2}2E_{p}2E_{p^{'}}}}\int{\dfrac{d^{3}k}
{(2\pi)^{3}2E_{k}}}\mid M\mid^{2}\nonumber\\
&\times(2\pi)^{3}\delta{(E_{p}+E_{p^{'}}-E_{k)}}\delta^{2}(p+p^{'}-k)\nonumber\\
&\times ((1-f_{q}(p_{z}))(1-f_{\bar{q}}(p^{'}_{z}))f_{g}(k)\nonumber\\
&\times -f_{q}(p_{z})f_{\bar{q}}(p^{'}_{z}))( 1+f_{g}(k))).
\end{align}
Here we are using LLL approximation. We are considering $p_{2}=0$ such that we will get $k_{z}=p_{z}+p^{'}_{z}$ In this process $p\equiv (p_{z},p_{2})$ this is due to the dimension reduction in presence of magnetic field. Matrix element is 
\begin{equation}
\mid M\mid^{2}= 8\pi \alpha_{eff} C_{2}e^{-\dfrac{k_{\perp}^{2}}{2\mid eB\mid}}\dfrac{k_{\perp}^{2}}{\mid k\mid^{2}}( E_{p}E_{p^{'}}+p_{z}p_{z^{'}}+m^{2}),
\end{equation}
where $\exp{\left( -\dfrac{k_{\perp}^{2}}{2\mid eB\mid}\right) }$ is the form factor. In this fundamental definition of 1 $\rightarrow$ 2 process collision integral we are using the detailed balance condition of quasi-parton distribution function. At equilibrium, the collision integral should be zero. This implies
\begin{align}\label{55}
&(1-f_{q}^{0}(p_{z}))(1-f_{\bar{q}}^{0}(p^{'}_{z}))f^{0}_{g}(k)\nonumber\\
&=f^{0}_{q}(p_{z})f^{0}_{\bar{q}}(p^{'}_{z}))( 1+f_{g}^{0}(k)),
\end{align}
$f_{(g,q)}^{0}$ is the equilibrium quasi-parton function. Using Eqs.~(\ref{36}),~(\ref{37}) we can write 
\begin{align}
&((1-f_{q}(p_{z}))(1-f_{\bar{q}}(p^{'}_{z}))f_{g}(k)\nonumber\\
&\times -f_{q}(p_{z})f_{\bar{q}}(p^{'}_{z}))( 1+f_{g}(k)))\nonumber\\
=&[(1-f_{q}^{0}(E_{p})+\beta f_{q}^{0}(E_{p})(1-f_{q}^{0}(E_{p}))\chi_{q})\nonumber\\
&\times (1-f_{q}^{0}(E_{p^{'}})+\beta f_{q}^{0}(E_{p^{'}})(1-f_{q}^{0}(E_{p^{'}}))\chi_{\bar{q}})\nonumber\\
&\times(f_{g}^{0}(E_{k})+\beta f_{g}^{0}(E_{k})(1+f_{g}^{0}(E_{k}))\chi_{g})]-\nonumber\\
&[(f_{q}^{0}(E_{p})+\beta f_{q}^{0}(E_{p})(1-f_{q}^{0}(E_{p}))\chi_{q})\nonumber\\
&\times(f_{q}^{0}(E_{p^{'}})+\beta f_{q}^{0}(E_{p^{'}})(1-f_{q}^{0}(E_{p^{'}}))\chi_{\bar{q}})\nonumber\\
&\times(1+f_{g}^{0}(E_{k})+\beta f_{g}^{0}(E_{k})(1+f_{g}^{0}(E_{k}))\chi_{g})].
\end{align}
Considering only the linear response term and from Eq.~(\ref{55}) we can show the detailed balance of quasi-parton distribution function as
\begin{align}
&((1-f_{q}(p_{z}))(1-f_{\bar{q}}(p^{'}_{z}))f_{g}(k)-\nonumber\\
&f_{q}(p_{z})f_{\bar{q}}(p^{'}_{z}))( 1+f_{g}(k)))\nonumber\\
&=\beta f^{0}_{q}(E_{p})f^{0}_{q}(E_{p^{'}})(1+f^{0}_{g}(E_{k}))\nonumber\\
&\times(\chi_{g}(k)-\chi_{\bar{q}}({p^{'}_{z}})-\chi_{q}(p_{z}))
\end{align}
Using this detailed balance of the quasi-parton distribution functions we have the collision integral as in Eq.~(\ref{39}).  

{}

\end{document}